\documentclass{PoS}

\title{Axial U(1) symmetry in the chiral symmetric phase of 2-flavor QCD at finite temperature}

\ShortTitle{Axial U(1) symmetry in the chiral symmetric phase of 2-flavor QCD at finite temperature}

\author{\speaker{Sinya Aoki} for JLQCD Collaboration\\
        Yukawa Institute for Theoretical Physics, Kyoto University, Kitashirakawa Oiwakecho, Sakyo-ku, Kyoto 606-8502, Japan, and\\
Center for Computational Sciences, University of Tsukuba, Tsukuba 305-8577, Japan\\
        E-mail: \email{saoki@yukawa.kyoto-u.ac.jp}}

\abstract{We discuss the fate of the axial U(1) symmetry in 2-flavor QCD at finite temperature, where the non-singlet chiral symmetry is restored. We first summarize the previous theoretical investigation on the relation between the eigenvalue density of the Dirac operator and the axial U(1) symmetry. We show that the eigenvalue density near the origin behaves as $\lambda^\gamma$ with $\gamma > 2$ in the chirally symmetric phase,  where $\lambda$ is an eigenvalue. This implies that the axial U(1) symmetry is partially restored, so that the low energy symmetry of the theory becomes  SU(2)$\otimes$ SU(2)$\otimes$ Z$_4$. 
Secondly, we report recent numerical investigations on this issue by lattice QCD simulations with lattice chiral fermions such as Overlap or improved domain-wall fermions. Our preliminary results indicate that the eigenvalue density seems to have a gap at the origin just above $T_c$, the temperature of the chiral symmetry restoration, which implies the axial U(1) symmetry is effectively restored at high temperature. We also point out an importance of the exact lattice chiral symmetry to obtain correct results on this issue.  }

\FullConference{The 8th International Workshop on Chiral Dynamics, CD2015 ***\\
		29 June 2015 - 03 July 2015\\
		Pisa,Italy}

\begin{document}

\section{Introduction}
The non-singlet chiral symmetry of QCD, SU($N_f)_L\otimes$ SU($N_f)_R$, which is spontaneously broken down to  SU($N_f)_V$ at low temperature, is expected to be restored above the critical temperature $T_c$.
The singlet axial U(1) symmetry, on the other hand, is broken explicitly by the anomaly at low temperature, and the fate of this symmetry above $T_c$ still remains an open issue.
Whether the axial U(1) (U(1)$_A$) symmetry is effectively restored or not near $T_c$ has a phenomenological importance\cite{Pisarski:1983ms,Pelissetto:2013hqa}. 

The Banks-Cahser relation\cite{Banks:1979yr} gives
$\displaystyle
\lim_{m\rightarrow 0} \langle \bar\psi \psi \rangle = \pi \rho(0),
$
 which says that $\rho(0)=0$ if the non-singlet chiral symmetry is restored at $T> T_c$,
while Ref.~\cite{Cohen:1996ng} pointed out that 
U(1)$_A$ symmetry is also restored if  $\rho(\lambda)$ has a gap near the origin such that
$\rho(\lambda) = 0$ for $^\forall\lambda\le \lambda_0\not=0$.

Other quantities sensitive to the U(1)$_A$ symmetry  are susceptibilities of scalar and pseudo-scalar operators, defined by $\chi^\sigma -\chi^\eta$ and $\chi^\pi -\chi^\delta$ for $N_f=2$, where
\begin{equation}
\chi^{X} =\lim_{V\rightarrow\infty}\frac{1}{V}\int d^4x\, \langle M^{X}(x) M^{X}(0)\rangle, \quad
M^{X} (x) = \bar\psi (x) (\Gamma_X\otimes T_X)  \psi(x), \quad V=\int d^4 x
\end{equation}
for $\Gamma_\sigma\otimes T_\sigma = 1\otimes 1$ (scalar singlet),  $\Gamma_\eta\otimes T_\eta = \gamma_5\otimes 1$(pseudo-scalar singlet), $\Gamma_\pi\otimes T_\pi = \gamma_5\otimes \tau^a$(pseudo-scalar triplet) and $\Gamma_\delta\otimes T_\delta = 1\otimes \tau^a$(scalar triplet).
If the U(1)$_A$ symmetry is restored, $\chi^\sigma -\chi^\eta$ and $\chi^\pi -\chi^\delta$ vanish in the chiral limit.

In this report, we first derive constraints on  the eigenvalue density of the Dirac operator using purely analytic method, and address the fate of U(1)$_A$ symmetry at $T\ge T_c$.
We then present our preliminary results  of recent numerical investigations on the fate of U(1)$_A$ symmetry, considering both eigenvalue density and U(1)$_A$ susceptibility.

\section{Theoretical investigation}
We here present the previous theoretical investigation in Ref.~\cite{Aoki:2012yj} on the fate of U(1)$_A$ symmetry in the chiral symmetric phase.

\subsection{Set up}
\begin{figure}[bth]
\vskip -0.8cm
\begin{center}
\scalebox{0.25}{
\includegraphics
{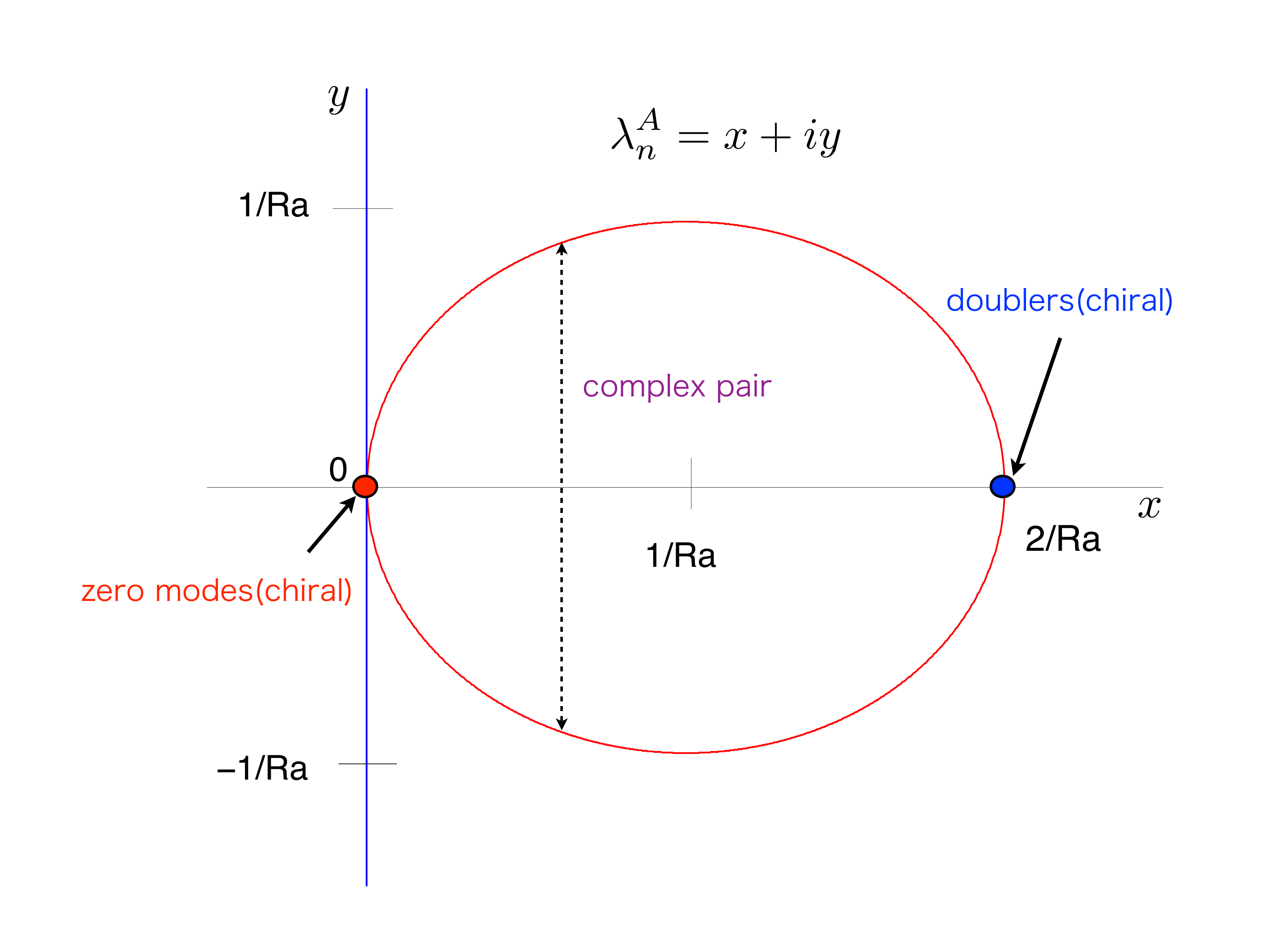}}\hfill
\end{center}
\vskip -0.5cm
\caption{Eigenvalues $\lambda_n^A=x+iy$ of the overlap Dirac operator $D(A)$. }
\label{fig:circle}
\end{figure}

We mainly consider 2-flavor QCD, using the lattice regularization, in order to avoid ambiguities associated with divergences in the continuum theory. 
For quarks, we employ the overlap fermion formulation, which poses the exact lattice ``chiral" symmetry
compatible with U(1)$_A$ anomaly through the Ginsparg-Wilson (GW) relation\cite{Ginsparg:1981bj} that
$D(A) \gamma_5 + \gamma_5 D(A) = a D(A) R \gamma_5 D(A)$,
 where $D(A)$ is the overlap Dirac operator of the massless theory for a given gauge configuration $A$, $a$ is the lattice spacing, and $R$ is a real parameter.
The GW relation gives
$\lambda_n^A + \bar\lambda_n^A = a R \bar\lambda_n^A \lambda_n^A$,  
where $\lambda_n^A$ is  an eigenvalue of $D(A)$ satisfying $D(A)  \phi_n^A = \lambda_n^A\phi_n^A$
for a eigenfunction $\phi_n^A$, while its complex conjugate $\lambda_n^A$ also becomes an eigenvalue as $D(A)  \gamma_5\phi_n^A = \bar\lambda_n^A\gamma_5\phi_n^A$ via the GW relation.
Note that real eigenvalues must be either $\lambda_0^A=0$ (zero mode) or
$\lambda_n^A = 2/Ra$ (doubler mode), both of which can be made chiral. 
Eigenvalues $\lambda_n^A = x+ i y$ forms a circle as shown in Fig.~\ref{fig:circle} 

The quark propagator can be expressed also in terms of these eigenvalues and eigenfunctions as
\begin{eqnarray}
D^{-1}(A)(x,y) &=&\sum_n\left[ \frac{\phi_n^A(x)(\phi_n^A)^\dagger(y)}{f_m \lambda_n^A +m}
+ \frac{\gamma_5\phi_n^A(x)(\phi_n^A)^\dagger(y)\gamma_5}{f_m \bar\lambda_n^A +m}
\right] \nonumber \\
&+&\sum_{k=1}^{N^A_{R+L}}\frac{1}{m}\phi_k^A(x)(\phi_k^A)^\dagger(y)+\sum_{K=1}^{N_D^A}\frac{Ra}{2}\phi_K^A(x)(\phi_K-A)^\dagger(y)
\label{eq:prop}
\end{eqnarray}
with $f_m= 1 -Rma/2$, where $m$ is the quark mass, $\phi_n^A$ represents a bulk mode, which is non-chiral due to the complex pair, and 
$\phi_k^A$ is a chiral zero mode, while $\phi_K^A$ is a chiral doubler mode.
The total number of chiral zero (doubler) modes is given by $N_{R+L}^A$ ($N_D^A$).
A measure for a gauge filed is then written as
\begin{eqnarray}
P_m(A) &=& e^{-S_{\rm YM}(A)} m^{N_f N_{R+L}^A}\left(\Lambda_R \right)^{N_f N_D^A}
\prod_{{\rm Im}\, \lambda_n^A > 0} \left( Z_m^2 \vert \lambda_n^A\vert^2 + m^2\right)^{N_f},
\end{eqnarray}
where $S_{\rm YM}$ is the gauge action, $\Lambda_R= 2/(Ra)$ and  $Z_m^2 =1-m^2/\Lambda_R^2$.
It is clear that $P_m$ is positive definite and an even function of non-zero $m$ for even $N_f$.
In this report, we take $N_f=2$.

A restoration of non-singlet chiral symmetry is compactly expressed for susceptibilities as
\begin{equation}
\lim_{m\rightarrow 0} \langle \delta^aO_{n_1,n_2,n_3,n_4}\rangle_m = 0,\quad
O_{n_1,n_2,n_3,n_4} =  (P^a)^{n_1} (S^a)^{n_2}(P^0)^{n_3} (S^0)^{n_4},
\label{eq:WTI}
\end{equation}
where $S^{a,0} =\int d^4 x\, S^{a,0}(x)$ and $P^{a,0} =\int d^4 x\, P^{a,0}(x)$ with scalar and pseudo-scalar densities, $S^{a,0}(x)$ and $P^{a,0}(x)$, respectively.
Here $\delta^a$ is the infinitesimal non-singlet chiral rotation, under which  
scalar and pseudo-scalar densities transform for $N_f=2$ as
\begin{eqnarray}
\delta^a S^b = 2\delta^{ab} P^0, \quad \delta^a P^b = - 2\delta^{ab} S^0,
\quad \delta^a S^0 = 2 P^a, \quad \delta^a P^0 = - 2 S^a .
\end{eqnarray}

\subsection{Assumptions}
In addition to  our basic assumption that non-singlet chiral symmetry is restored at $T\ge T_c$,
we make following two assumptions in our analysis.

If $O(A)$ is $m$-independent, we assume 
\begin{eqnarray}
\langle O(A) \rangle_m :=\frac{1}{Z}\int {\cal D}A\, P_m(A) O(A) = f(m^2),
\end{eqnarray}
where $f(x)$ is analytic at $x=0$. Note that this assumption does not hold if the chiral symmetry is spontaneously broken. For example,
\begin{equation}
\lim_{V\rightarrow\infty} \frac{1}{V} \langle Q(A)^2\rangle_m = m \frac{\Sigma}{N_f} + O(m^2)
\end{equation}
at $T < T_c$, where $Q(A)$ is the topological charge and $\Sigma$ is the chiral condensate.

We also assume that the eigenvalue density of the overlap Dirac operator can be expanded as
\begin{eqnarray}
\rho^A(\lambda)&:=& \lim_{V\rightarrow\infty}\frac{1}{V}\sum_n \delta\left(\lambda -\sqrt{\bar\lambda_n^A\lambda_n^A}\right) =\sum_{k=0}^\infty \rho_k^A\frac{\lambda^k}{k!}, 
\end{eqnarray}
for a small $\lambda$. More precisely, configurations whose eigenvalue density can not be expanded at the origin are measure zero in the configuration space.

At non-zero lattice spacing, integrals of $\lambda^k$ over all eigenvalues are convergent, since all eigenvalues satisfy the upper bound such that $\vert \lambda_n^A\vert \le \Lambda_R$.

\subsection{Analysis}
In this subsection, we present some examples of our analysis.

We first consider a constraint for $O_{1,0,0,0}$ as
\begin{eqnarray}
\lim_{m\rightarrow 0}\lim_{V\rightarrow\infty} \frac{\langle \delta^a P_a\rangle_m}{2V}  &=&
\lim_{m\rightarrow 0}\lim_{V\rightarrow\infty} \frac{\langle  - S_0\rangle_m}{V} = 0 .
\label{eq:1pt}
\end{eqnarray}
Using eq.~(\ref{eq:prop}), we have
\begin{eqnarray}
\lim_{V\rightarrow\infty} \frac{\langle - S_0\rangle_m}{V} &=& \lim_{V\rightarrow\infty} \frac{N_f}{mV}
\langle N_{R+L}^A\rangle_m + N_f \langle I_1 \rangle_m, 
\end{eqnarray}
where
\begin{equation}
I_1 = \int_0^{\Lambda_R} d\lambda\, \rho^A(\lambda) g_0(\lambda)\frac{2m}{Z_m^2\lambda^2+m^2}  = \pi \rho_0^A + O(m)
\end{equation}
with $g_0(x^2)= 1-x^2/\Lambda_R^2$. Here a source of $m$ singularities is  $2m/(Z_m^2\lambda^2+m^2)$, which comes from $D^{-1}(A)$, not from $\rho^A$.
Eq.~(\ref{eq:1pt}) implies
\begin{eqnarray}
\langle \rho_0^A\rangle_m &=& O(m^2), \quad \lim_{V\rightarrow\infty} \frac{\langle N_{R+L}^A\rangle_m}{V} =O(m^2) .
\end{eqnarray}

From  eq.~(\ref{eq:WTI}) for $O_{1,0,0,N-1}$ with an arbitrary $N$, we obtain 
\begin{eqnarray}
0&=& -\frac{1}{V^N} \langle S_0^N \rangle_m = (-1)^{N+1}N_f^N\left\langle
\left(\frac{N_{R+L}^A}{mV} + I_1\right)^N\right\rangle +\left(\frac{1}{V}\right)
\end{eqnarray}
in the large volume, which leads to
\begin{equation}
\lim_{V\rightarrow\infty}\frac{\langle N_{R+L}^V\rangle_m}{V} = 0 .
\end{equation}

We next consider $O_{0,1,1,0}$, which gives
\begin{eqnarray}
0 &=& \lim_{m\rightarrow 0}\lim_{V\rightarrow \infty}\chi^{\eta-\delta}=
 -\lim_{m\rightarrow 0}\lim_{V\rightarrow \infty} \frac{N_f^2\langle Q^A\rangle_m}{m^2V^2}
 +N_f \lim_{m\rightarrow 0}\left\langle \left(\frac{I_1}{m} + I_2\right)\right\rangle_m,
\end{eqnarray}
where
\begin{eqnarray}
I_2 &=& 2\int_0^{\Lambda_R}d\lambda\, \rho^A(\lambda)\frac{m^2g_0^2(\lambda^2) -\lambda^2g_0(\lambda^2)}{(Z_m^2 \lambda^2+m^2)^2} \rightarrow
\frac{I_1}{m} + I_2 = \frac{\pi}{m}\rho_0^A +2\rho_1^A +O(m) ,
\end{eqnarray}
and $Q^A = N_R^A - N_L^A$ is the index of the overlap Dirac operator, which gives a definition of the topological charge for a given gauge configuration $A$.
We then obtain
\begin{eqnarray}
\lim_{m\rightarrow 0}\lim_{V\rightarrow \infty} 
\frac{N_f\langle Q^A\rangle_m}{m^2V^2}
&=& 2 \lim_{m\rightarrow 0} \langle \rho_1^A\rangle_m .
\end{eqnarray}

\subsection{Final results}
Repeating similar analysis for higher susceptibilities, we finally obtain
\begin{eqnarray}
\lim_{m\rightarrow 0} \langle \rho^A(\lambda)\rangle_m &=& \lim_{m\rightarrow 0} \langle \rho^A\rangle_m \frac{\lambda^3}{3!} + O(\lambda^4).
\label{eq:results}
\end{eqnarray}
It is noted that we should not obtain no more constraints on higher $\langle \rho_n^A\rangle_m$ from general considerations, since $\displaystyle\lim_{m\rightarrow 0} \langle \rho_3^A\rangle_m\not=0$ even for a free theory, which poses both non-singlet and singlet  chiral symmetries.
In addition to the above result, we have
\begin{eqnarray}
\langle \rho_0^A\rangle_m &=& 0, \quad
\lim_{V\rightarrow\infty}\frac{1}{V^k} \langle (N_{R+L}^A)^k \rangle_m = 0, \quad
\lim_{V\rightarrow\infty}\frac{1}{V^k} \langle (Q^A)^{2k} \rangle_m = 0
\end{eqnarray}
for small but non-zero $m$.

\subsection{Consequences}
The result in the previous subsection leads to the constraint for the singlet susceptibility at $T \ge T_c$ as
\begin{equation}
\lim_{m\rightarrow 0}\chi^{\pi-\eta} =
\lim_{m\rightarrow 0}\lim_{V\rightarrow\infty} \frac{N_f^2 \langle (Q^A)^2\rangle_m}{m^2 V} = 0 ,
\end{equation}
which implies that screening lengths of $\pi$ and $\eta$ are equal: $\xi_\pi = \xi_\eta$. 
This result,  however, is necessary but not sufficient condition for the restoration of the U(1)$_A$ symmetry.

The overlap fermion satisfies the anomalous  Ward-Takahashi identities for the singlet  as
\begin{eqnarray}
\langle J^0 O + \delta^0 O \rangle_m = O(m),
\label{eq:AWTI}
\end{eqnarray}
where $J^0$ is the measure term representing the anomaly and $\delta^0$ is the flavor singlet chiral transformation. For $O_{n_1,n_2,n_3,n_4}$, we can show 
\begin{eqnarray}
\lim_{V\rightarrow\infty} \frac{1}{V^k} \langle J^0 O\rangle_m 
&=& \lim_{V\rightarrow\infty} \left\langle \frac{(Q^A)^2}{mV} \times O(V^0)\right\rangle_m = 0,
\end{eqnarray}
 where $k$ is the smallest integer which makes the $V\rightarrow\infty$ limit finite.
 Combining this with the anomalous Ward-Takahashi identities (\ref{eq:AWTI}), we conclude that
 \begin{eqnarray}
\lim_{m\rightarrow 0} \lim_{V\rightarrow\infty} \frac{1}{V^k} \langle \delta^0 O\rangle_m = 0, 
\end{eqnarray}
which means that the violation of U(1)$_A$ symmetry by the anomaly becomes invisible for these bulk susceptibilities at $T \ge T_c$.
Thus the low energy effective theory for pseudo-scalar mesons in 3-dimensions poses
SU(2)$_L \otimes$ SU(2)$_R \otimes$ Z$_4$ symmetry at $T\ge T_c$, rather than
SU(2)$_L \otimes$ SU(2)$_R \otimes$ U(1)$_A$ expected for the restration of the U(1)$_A$ symmetry.
 
There are a few remarks. To obtain results in this section, the following conditions are important:
the large volume limit as $V\rightarrow \infty$, the chiral limit as $m\rightarrow 0$ and the lattice chiral symmetry realized by the GW relation that $D\gamma_5 +\gamma_5 D = a D R \gamma_5 D$.
A lack of one of these conditions in numerical simulations may easily lead to incorrect conclusions.

We can generalize our analysis in the case of the fractional power for the eigenvalue density such that
$
\rho^A(\lambda) \simeq c_A \lambda^\gamma.
$
 If the non-singlet chiral symmetry is restored, we can derive the condition that $\gamma > 2$, which is consistent with eq.~(\ref{eq:results}).
 
 \section{Recent numerical results}
 \subsection{Status of recent numerical simulations}
 \begin{table}[bt]
\vskip -0.2cm
\begin{center}
\begin{tabular}{|c|c|c|c|c|c|}
\hline 
Group & Quark & Size & $\rho(\lambda)$ &  U(1)$_A$ mesons  &U(1)$_A$ \\
\hline\hline
JLQCD13\cite{Cossu:2013uua} & OV with fixed $Q$ & 2 fm & gap & degenerate & restored \\
\hline
TWQCD13\cite{Chiu:2013wwa} &optimal DW & 3 fm & no gap & degenerate & restored \\
\hline
HotQCD12\cite{Bazavov:2012qja} &DW & 2,3 fm & no gap &  no & violated \\
LLNL/RBC13\cite{Buchoff:2013nra} & DW & 4 fm & no gap & no & violated \\
\hline
HotQCD14\cite{Bhattacharya:2014ara} & M\"obius DW & 4,11 fm & N/A & no & violated \\
\hline
Dick {\it et al.}15\cite{Dick:2015twa} &OV  on HISQ & 3,4 fm & no gap &  no & violated \\
\hline
\end{tabular}
\caption{Summary of recent investigations on U(1)$_A$ symmetry at $T > T_c$.
Results on the gap of the eigenvalue density $\rho(\lambda)$ and the degeneracy of U(1)$_A$ meson correlation functions are given in 4th and 5th columns, respectively.
The conclusion on  U(1)$_A$ symmetry just above $T_c$ is given in the last column. }
\end{center}
\vskip -0.5cm
\label{tab:summary}
\end{table}
Although many numerical investigations in lattice QCD have been attempted from the end of the last century, in order to answer the question whether U(1)$_A$ symmetry is restored or not in the chirally symmetric phase at finite temperature, so far no definite conclusion is obtained.
Situation gets worse recently, as the eigenvalue density  of lattice Dirac operators has been calculated numerically. Table~\ref{tab:summary} shows a summary of recent investigations.

While the JLQCD collaboration\cite{Cossu:2013uua}  and TWQCD collaboration\cite{Chiu:2013wwa}
have concluded that U(1)$_A$ symmetry is restored just above $T_c$, HotQCD collaboration\cite{Bazavov:2012qja,Bhattacharya:2014ara} and LLNL/RBC collaboration\cite{Buchoff:2013nra} have reported that
U(1)$_A$ symmetry is still violated just above $T_c$.   All these collaborations have employed the lattice chiral quarks such as overlap (OV) and domain-wall (DW) formulations  with various improvements in their numerical simulations, which pose a good chiral symmetry even at finite lattice spacing. 
In addition, Dick {\it et al.} has employed the valence overlap quarks on configurations generated by highly improved staggered quark(HISQ) action, and also concluded the violation of  U(1)$_A$ symmetry.

What causes these differences ? Possible sources which cause differences are lattice volumes, quark masses and quark actions.  
One may naively think that the difference between the overlap quark employed,
which  satisfies an exact GW relation, and the (improved) domain-wall  quark,  which has an approximated one,  is numerically very tiny.
We have recently found, however, that such a tiny difference can produce large difference in the eigenvalue density and the U(1)$_A$ susceptibility.
In this section, we report our preliminary results on these two points. 
 
\subsection{Eigenvalue density, partially quenching and reweighting} 
Tomiya {\it et al.} for JLQCD collaboration\cite{Tomiya:2014mma} have attempted the following analysis.
They first generated gauge configurations in 2-flavor lattice QCD with an improved DW quarks, which has very small violation of GW relation\cite{Hashimoto:2014gta}.
Then eigenvalue density $\rho(\lambda)$ has been evaluated in three different ways.\\
(0)They calculate $\rho(\lambda)$ of the improved DW operator on configurations generated by the same improved DW quark.
This is called the original eigenvalue density.\\
(1) They calculate  $\rho(\lambda)$ of the overlap operator on these configurations.
Since the valence and sea quarks are different, this is called the partially quenched  eigenvalue density. \\
(2) The reweighing factor from the improved DW to overlap quarks is calculated in order to obtain the full overlap eigenvalue distribution from the partially quenched one. This is called the full overlap eigenvalue density. 

\begin{figure}[bt]
\vskip -0.9cm
\begin{center}
\scalebox{0.25}{
\includegraphics[width=4.0\textwidth]
{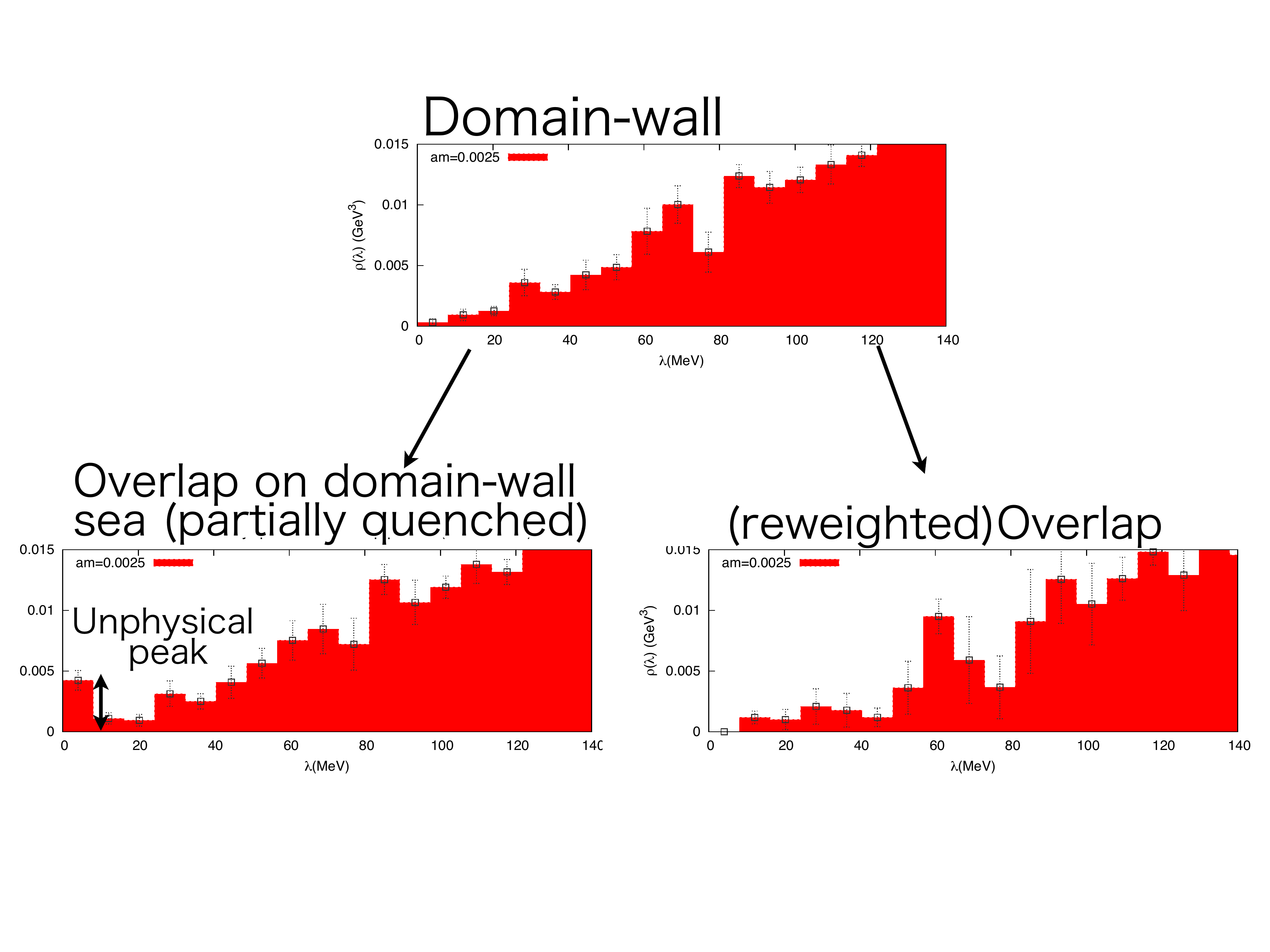}}
\end{center}
\vskip -2.5cm
\caption{Eigenvalue  density $\rho(\lambda)$ for three different cases at $T\simeq 190 {\rm MeV} \simeq 1.05 T_c$ and $L\simeq3$ fm. (Upper) DW operator on DW configurations. (Lower-Left) OV operator on DW configurations. (Lower-Right) OV operator on OV configurations by reweighting.
The horizontal axis in each figure is the eigenvalue $\lambda$ in unit of MeV, while the vertical axis is the density $\rho(\lambda)$ in unit of GeV$^3$, normalized by the spatial volume.   }
\label{fig:density}
\vskip -0.25cm
\end{figure}

Fig.~\ref{fig:density} shows the latest comparison of these 3 different $\rho(\lambda)$, reported in LATTICE 2015\cite{Tomiya:2015}. 
These eigenvalue densities  are obtained at $T\simeq 190$ MeV $\simeq 1.05 T_c$, just above $T_c\simeq 180$ MeV, on $L\simeq 3$ fm, where $L$ is the spatial lattice size. The bare quark mass is $m a = 0.0025$ in lattice unit, which corresponds to $m \simeq 6$ MeV  at $T\simeq 190$ MeV.  

The upper figure is the original eigenvalue density (DW operator on DW configurations), which shows strong suppression of small eigenvalues, suggesting the restoration of U(1)$_A$ symmetry. More detailed analyses including the infinite volume limit and the chiral limit, however, are needed for the definite conclusion.
The lower-left figure is the partially quenched eigenvalue density (OV operator on DW configurations), which
shows a peak at smaller eigenvalue near the origin, in contrast to the original density. 
This peak must be an artifact of partially quenching, therefore unphysical,  since small eigenvalues are suppressed in the original density.  
In general, configurations which give small eigenvalues of the OV operator are not suppressed by
the determinant of the DW quark, which is insensitive to such small eigenvalues.
If configurations were generated by the OV quarks, appearance of small eigenvalues of the OV operator
would be highly suppressed.
This expectation is indeed the case, as seen in the lower-right figure, which shows the full overlap eigenvalue density (OV operator on OV configurations by reweighting).
The peak presented at the origin for the partially quenched case completely disappears. 
Furthermore, small eigenvalues are more strongly suppressed in the full OV than in the original DW, suggesting
an appearance of the gap at the origin. 
  
The above results, though still preliminary, give us an important lesson that an exact lattice chiral symmetry is essential to obtain the correct conclusion. A tiny violation of the chiral symmetry may destroy the theoretically expected behavior of the density.  In particular, partially quenched calculations are dangerous: They are sometimes worse than original calculations with less lattice chiral symmetry, since a mismatch between two Dirac operators artificially enhances appearance of small eigenvalues.

\subsection{U(1)$_A$ susceptibility}
In ref.~\cite{Cossu:2014aua}, Cossu {\it et al.} consider the U(1)$_A$ susceptibility $\Delta$, defined by
\begin{eqnarray}
\Delta &\equiv& \chi^\pi - \chi^\delta = \frac{2 \langle N_{R+L}^A \rangle_m }{Vm^2} + 
\mbox{ Bulk non-zero modes} .
\end{eqnarray}
We here discuss  the update of the result  in ref.~\cite{Cossu:2015,Cossu:2015a}, where same configurations in the previous subsection are employed. 

\begin{figure}[bt]
\vskip -0.3cm
\begin{center}
\includegraphics[width=0.32\textwidth]{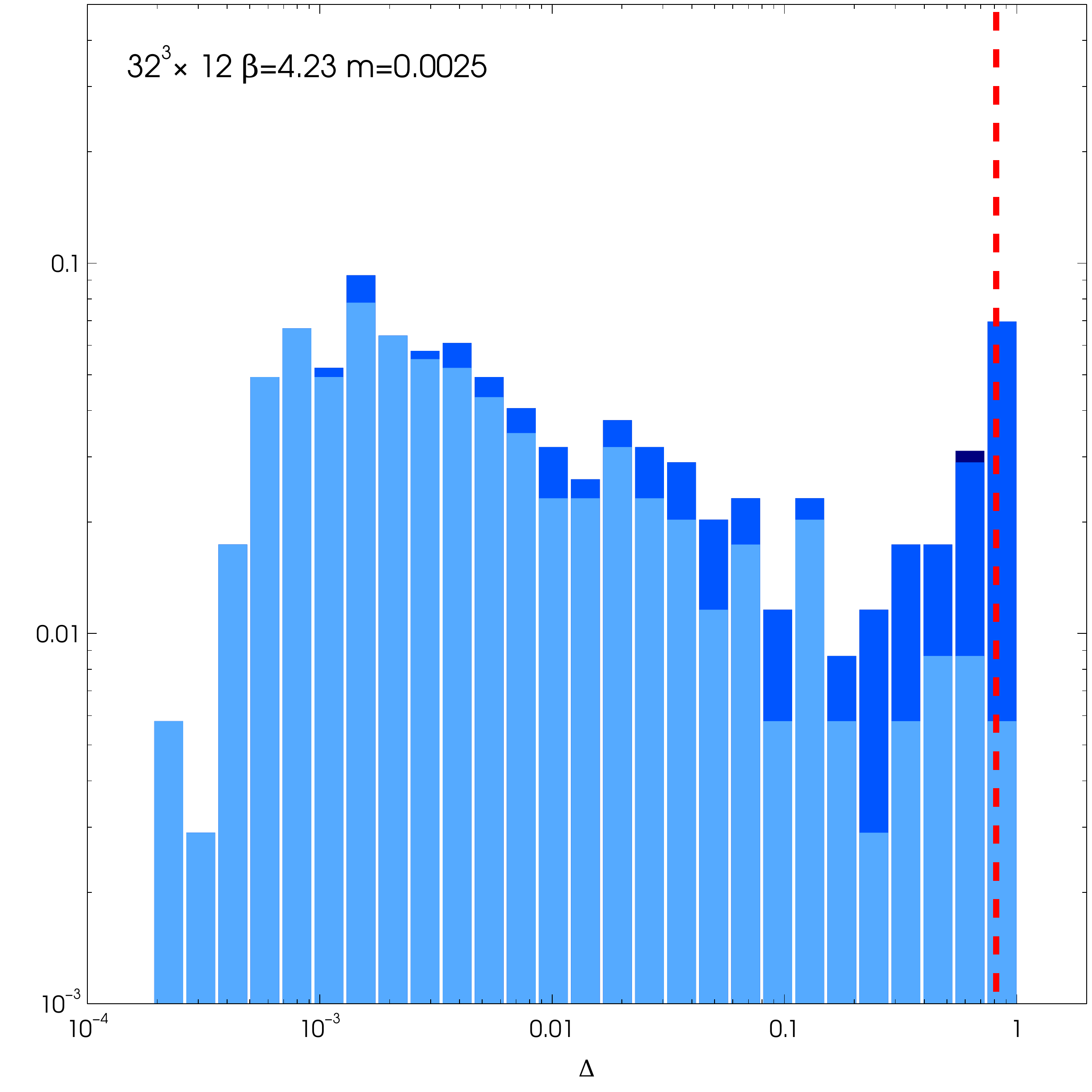}
\includegraphics[width=0.32\textwidth]{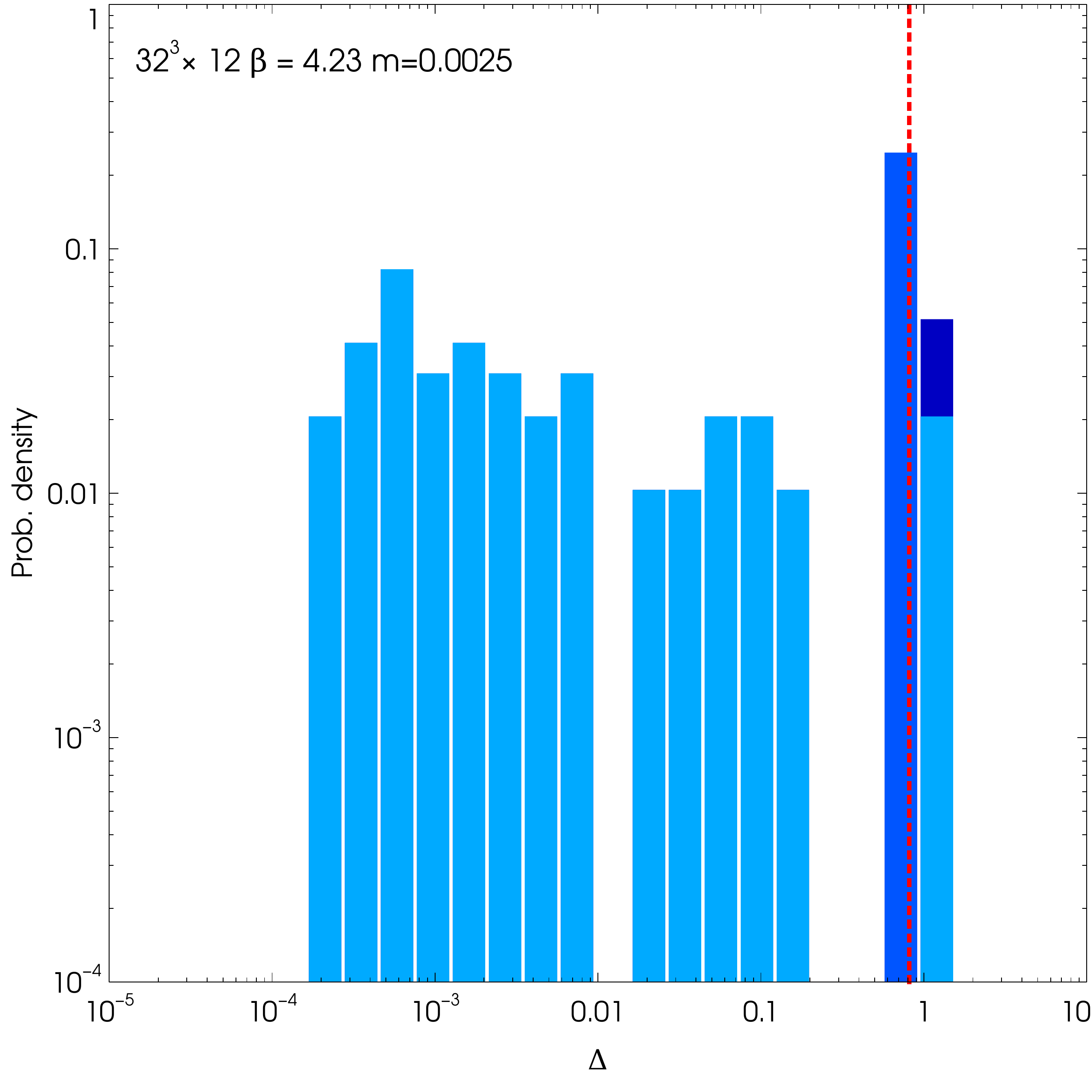}
\includegraphics[width=0.32\textwidth]{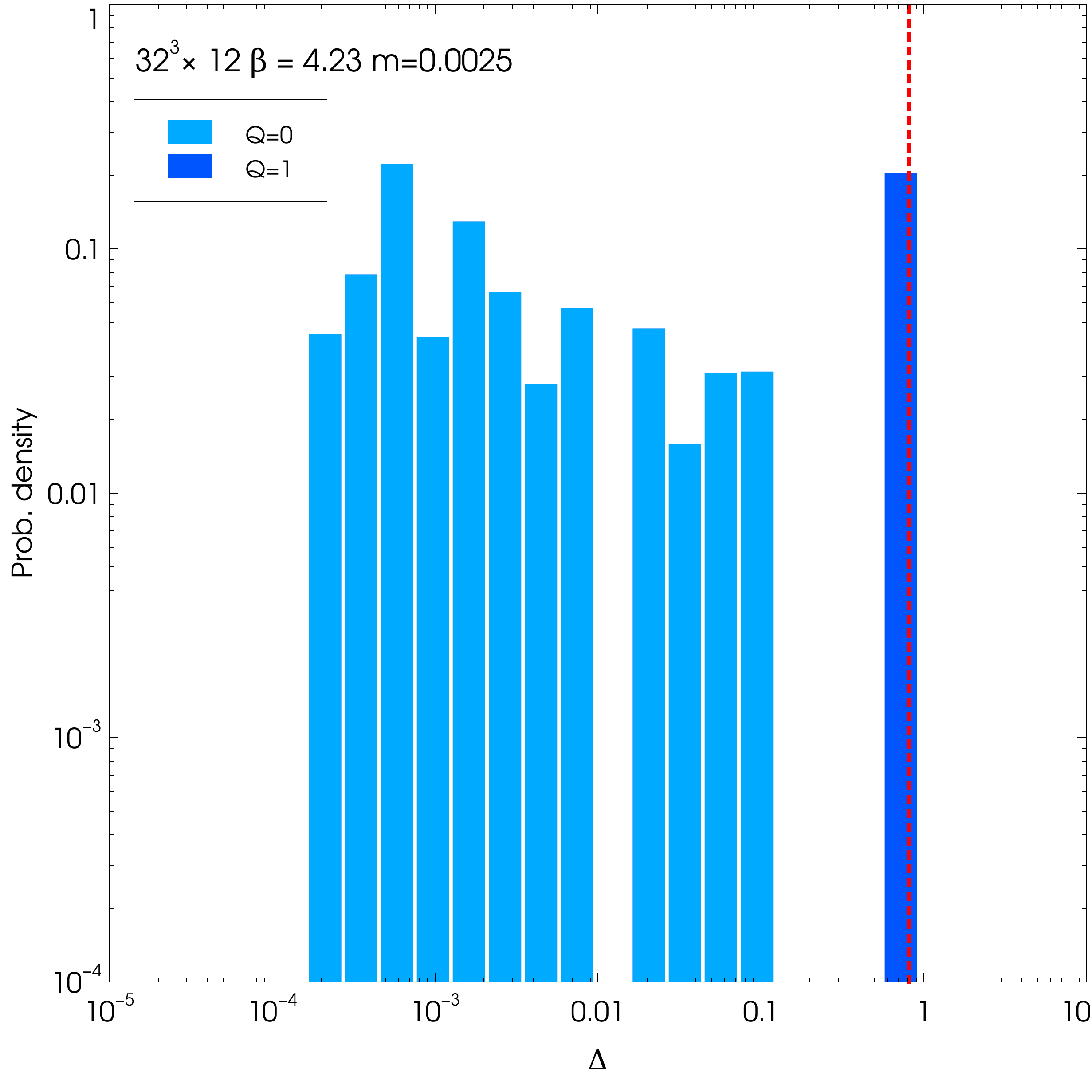}
\end{center}
\vskip -0.5cm
\caption{Distribution of $\Delta$ for three different cases at $T\simeq 190 {\rm MeV} \simeq 1.05 T_c$ and $L\simeq3$ fm. (Left) DW quarks on DW configurations. (Center) OV quarks on DW configurations. (Right) OV quarks on OV configurations by reweighting.
The horizontal axis shows $\Delta$ in the log scale.   }
\label{fig:delta}
\end{figure}  
In Fig.~\ref{fig:delta}, the distribution of $\Delta$ is plotted for three different cases as in the previous subsection. 
In the left figure, $\Delta$ is calculated the DW propagator on DW configurations. 
The vertical dashed  red line around $\Delta =1$ corresponds to the zero mode contribution, $ \displaystyle\frac{2 N_{R+L}^A  }{Vm^2}$ at this quark mass and volume. 
The light blue part of the histogram represents the contribution from configurations with $N_{R+L}^A = Q^A=0$,
while the dark blue corresponds to the one with $N_{R+L}^A = Q^A=1$. 
On this volume, the zero mode contribution would dominate in $\Delta$ if the GW relation were exact,
so that the dark blue would appear on the dashed red line while the light blue would give much smaller $\Delta$.  
Therefore, the spread of both light blue and dark blue over the large range of $\Delta$ in the log scale 
indicates that a tiny violation of the GW relation of the improved DW quark gives a big impact on $\Delta$, which is an indicator of the U(1)$_A$ symmetry violation.

The right figure shows the histogram of $\Delta$ for the OV propagator on OV configurations by reweighting, where $Q^A=1$ contributions only appear on the dashed red line while $Q^A=0$ contributions
give an order of magnitude smaller values. Since zero mode contributions are expected to vanish in the infinite volume limit, $\Delta$ becomes much smaller in that limit.
The central one corresponds to the partially quenched calculation by the OV propagator on DW configurations.
Although $Q^A=1$ contributions are evaluated correctly and a majority of $Q^A=0$ contributions is small,
some $Q^A=0$ configurations give large contributions comparable to those with $Q^A=1$.
As in the previous subjection, partially quenched results show accumulation of unphysical near zero modes,
which produce a large U(1)$_A$ violation.  

Results in this subsection also support the conclusion in the previous subsection. 

\section{Conclusion}
Both analytic and numerical investigations reported here indicate that U(1)$_A$ symmetry is effectively restored, so that the low energy symmetry of 2-flavor QCD above $T_c$ is either SU(2)$_L \otimes$ SU(2)$_R \otimes$ Z$_4$ ($\rho(\lambda)$ is gapless) or SU(2)$_L \otimes$ SU(2)$_R \otimes$ U(1)$_A$ ($\rho(\lambda)$ has a gap). 
The conformal bootstrap method\cite{Nakayama:2014sba} confirms the perturbative prediction\cite{Pelissetto:2013hqa} that 
the 2nd order chiral phase transition is possible for either case,
though its universality class is different from the conventional SU(2)$_L \otimes$ SU(2)$_R\simeq$ O(4) chiral transition.

\begin{acknowledgments}
The author would like to thank Drs. Guido Cossu and Akio Tomiya for providing  their latest results.
 This work is supported in part by Grant-in-Aid for Scientific Research (B) 2528704,  MEXT SPIRE (Strategic Program for Innovative REsearch) and JICFuS.
\end{acknowledgments}

\end{document}